\RequirePackage{fix-cm}
\documentclass[onecolumn,epjc3]{svjour3}  
\usepackage{tikz}
\usepackage{amsmath}
\usepackage{amssymb}
\usepackage{booktabs}
\usepackage{amsfonts}
\smartqed  
\RequirePackage{graphicx}
\journalname{Eur. Phys. J. C}
\begin{document}

\title{Emergent Gribov horizon kernel from replica symmetry breaking in Yang--Mills theories}




\author{Rodrigo Carmo Terin}
\institute{
King Juan Carlos University, Faculty of Experimental Sciences and Technology, Department of Applied Physics,\\
Av.\ del Alcalde de M\'ostoles, 28933 Madrid, Spain\\
\email{rodrigo.carmo@urjc.es}
}

\date{Received: date / Accepted: date}

\maketitle

\begin{abstract}
We show that, in the replica-broken sector of the Serreau--Tissier (ST) gauge fixing,
the expansion of the replica determinant in the regulator $\zeta$ induces a
nonlocal bilinear gluonic kernel with the same color and Lorentz structure as
the quadratic part of the BRST-invariant Gribov horizon functional.
This establishes an effective leading-order correspondence with the refined Gribov-Zwanziger (RGZ) horizon
sector, rather than a reconstruction of the full nonlinear functional
$H(A^h)$.
The induced scale satisfies
$\gamma_{\mathrm{ind}}^4\propto \zeta$ at leading order, up to
scheme-dependent normalization and higher-order corrections.
Depending on the replica phase, the ST sector yields either a local
Curci--Ferrari (CF) screening mass or an induced RGZ-type horizon kernel, avoiding
double counting of infrared scales.
\end{abstract}

\section{Introduction}
\label{sec:intro}

Infrared quantum chromodynamics (QCD) is notoriously difficult to describe
because gauge fixing in a nonabelian gauge theory is obstructed by Gribov copies
and by the breakdown of naive perturbation theory
\cite{YangMills1954,FaddeevPopov1967,Gribov:1977,Singer:1978}.
Two complementary continuum frameworks have been developed to address these
difficulties.
On the one hand, J.~Serreau, M.~Tissier and collaborators proposed a family of
Landau gauges in which all extrema \(U\) of the Landau functional
\[
f[A,U]=\int d^dx\,{\rm tr}\big[(A_\mu^U)^2\big]
\]
are averaged with a non-uniform weight proportional to
\[
\frac{\det(F[A,U]+\zeta\mathbf{1})}{|\det F[A,U]|}
\exp[-\beta f[A,U]]
\]
\cite{Serreau:2012cg,SerreauTissierTresmontant2015,Tissier_2018,Reinosa:2020skx}.
A replica trick, inspired by the Parisi--Sourlas construction
\cite{ParisiSourlas1979}, localizes this copy average into a perturbatively
renormalizable action containing nonlinear sigma model superfields.
In the replica-symmetric phase, this construction yields a massive
Faddeev--Popov/Curci--Ferrari (FP/CF) type infrared behavior for the gluon
two-point function.

On the other hand, the Gribov--Zwanziger (GZ) and RGZ approaches implement the restriction of the functional
integral to the first Gribov region by means of the horizon functional
\cite{Zwanziger:1989mf,Zwanziger:1990tn,Dudal:2008sp,Dudal:2011gd}.
In the BRST-invariant formulation, this functional is written in terms of the
transverse gauge-invariant field $A_\mu^h$,
\begin{equation}
H(A^h)
=
g^2\int d^dx\,d^dy\,
f^{abc}A_\mu^{h,b}(x)
\big[\mathcal M^{-1}(A^h)\big]^{ad}(x,y)
f^{dec}A_\mu^{h,e}(y),
\label{eq:intro-horizon}
\end{equation}
where
\[
\mathcal M(A^h)=-\partial_\mu D_\mu(A^h)
\]
is the FP operator evaluated on $A_\mu^h$.
After localization through Zwanziger auxiliary fields and refinement by
dimension-two condensates, the RGZ theory yields an infrared-finite
decoupling-type gluon propagator compatible with lattice results
\cite{Capri:2015nzw,Capri:2017bfd,Capri:2021pye,PhysRevLett.90.152001,Cucchieri:2008PRL,DudalOliveiraRoelfsSilva2020}.
In a recent work
\cite{terin2025unifiedviewpointgribovzwanzigerserreautissier}, we constructed a
unified gauge-fixing framework combining the ST copy-averaged construction and
the RGZ restriction to the first Gribov region. The basic idea was to average
over Gribov copies with the combined weight
\begin{equation}
\exp\big[-\beta f[A,U_i]-\gamma^4H(A^{h,U_i})\big]\,
\frac{s(i)\,\det(F[A,U_i]+\zeta\,\mathbf{1})}{|\det F[A,U_i]|}.
\label{eq:intro-unified-weight}
\end{equation}
in which $s(i)=\mathrm{sign}\det F[A,U_i]$. Gauge-invariant operators are
unchanged by the copy average, while the limits $\gamma\to0$ and
$(\beta\to\infty,\zeta\to0)$ recover, respectively, the ST and GZ/RGZ
sectors. Localizing the ST weight by the replica trick and the horizon
functional through Zwanziger fields gives a single local action in which the
replica and RGZ sectors couple through a common gauge field. In the
$A^h$-based formulation, the resulting theory preserves a nilpotent BRST
symmetry and is power-counting renormalizable
\cite{terin2025unifiedviewpointgribovzwanzigerserreautissier}.

The present paper addresses a different question. Our previous unified
construction provided a BRST-consistent local framework in which copy averaging
and horizon suppression can coexist algebraically. It did not, however,
explain whether any part of the horizon structure can be generated dynamically
from the ST replica sector itself. Here we analyze this issue by integrating
out the replica superfields and studying the determinant generated by the ST
sector.

The main result is that, in the replica-broken sector, the expansion of the
replica determinant in the regulator $\zeta$ induces a nonlocal gluonic
kernel. At leading order in $\zeta$ and at quadratic order in $A^h$, this
kernel has the same color and Lorentz structure as the quadratic part of the
BRST-invariant Gribov horizon functional. More precisely, if
\begin{equation}
H(A^h)=H_2(A^h)+O((A^h)^3),
\end{equation}
then the ST determinant generates an effective contribution proportional to
$H_2(A^h)$, up to local terms, normalization conventions and higher-order
corrections in $\zeta$. Thus, the result should be understood as an effective
leading-order correspondence at the level of the nonlocal horizon kernel, not
as a derivation of the full nonlinear functional $H(A^h)$ to all orders.

The phase structure of the replica sector is fundamental in this
interpretation, i.e., in the replica-symmetric phase, characterized by a positive
replica curvature $\hat\chi>0$, integrating out the nonlinear sigma fields
leads to a local CF-like screening mass. In contrast, in the replica-broken
sector, represented by the limiting branch $\hat\chi=0$, nonlocal determinant
fluctuations survive and generate a horizon-like kernel. The two mechanisms
therefore correspond to different infrared realizations of the same replica
construction.
This phase dependence also avoids double counting of infrared scales. The
local CF screening contribution and the induced RGZ-type horizon kernel are
not treated as two independent ST-generated effects active in the same phase.
Rather, the replica-symmetric branch gives a local screening mass, while the
replica-broken branch gives an induced nonlocal horizon kernel. This separation
is implemented in the effective propagator through a phase factor
$\Xi_{\rm rep}$, introduced below.
The induced horizon scale is parametrized as
\begin{equation}
\gamma_{\mathrm{ind},R}^{4}
=
\kappa_R(d,N,\mu)\,\mu^2\,\zeta_R
+
O(\zeta_R^2),
\label{eq:intro-gamma-ind}
\end{equation}
where $\mu$ is the renormalization scale and the subscript $R$ denotes
renormalized quantities. The coefficient $\kappa_R(d,N,\mu)$ is defined by
projecting the induced nonlocal bilinear kernel onto the transverse
RGZ-horizon structure. It is therefore scheme dependent and not a universal
number. Its physical sign is fixed by requiring the induced contribution to
enter the quadratic action with the same sign as the usual RGZ horizon term in
the horizon-dominated branch.

The purpose of our present work is therefore not to replace the RGZ horizon
condition or to claim a full dynamical equivalence between the ST and RGZ
prescriptions. Instead, we show that the ST replica determinant contains, in a
well-defined infrared branch, the leading nonlocal structure responsible for
an RGZ-type horizon contribution. If an explicit RGZ horizon term is also
included, the induced contribution should be viewed as part of the effective
horizon strength,
\[
\gamma_{\rm eff}^4=\gamma^4+\gamma_{\rm ind}^4,
\]
with the usual RGZ horizon condition applying to the effective combination.
This construction suggests several possible tests. In lattice simulations, one
may compare Landau-gauge ensembles generated with different copy-selection or
copy-weighted prescriptions and study the corresponding infrared curvature of
the gluon propagator, the zero-momentum value $D_T(0)$, and RGZ-like complex
pole fits. In continuum approaches, the same mechanism could be probed through
Dyson--Schwinger equations, functional renormalization group equations, or
spectral reconstructions using an ST-induced horizon-like kernel.
We also discuss a statistical analogy between replica averaging and spin
systems, which provides a natural language for the competition between
screening and horizon suppression and suggests connections with modern
energy-based models
\cite{Hopfield1982,AckleyHintonSejnowski1985,TehWellingHinton2003}.
This analogy is only interpretative; the technical result of the present paper
is the explicit identification of the induced bilinear horizon kernel.

This paper is organized as follows. In Sec.~\ref{sec:review}, we briefly review
the unified ST--RGZ gauge fixing and recall how copy averaging and horizon
suppression can be inserted into a single local BRST-invariant action.
In Sec.~\ref{sec:replica-eff}, we integrate out the replica nonlinear sigma
superfields and derive the determinant expansion that generates the nonlocal
gluonic kernel. In Sec.~\ref{sec:horizon-functional}, we match the induced
bilinear kernel with the quadratic part of the BRST-invariant Gribov horizon
functional and define the induced scale $\gamma_{\mathrm{ind}}$.
In Sec.~\ref{sec:limits-interpretation}, we clarify the phase dependence of the
construction and the relation between $\gamma_{\mathrm{ind}}$ and the usual
RGZ horizon parameter. In Sec.~\ref{sec:gluon-propagator}, we compute the
corresponding tree-level gluon propagator and show explicitly how double
counting is avoided. The main implications, possible tests, and open problems
are discussed in Sec.~\ref{sec:discussion}. Appendix~\ref{sec:susy-horizon}
summarizes the superspace derivation and explains how the relevant supertrace
reduces to the ordinary trace over $\mathcal M^{-1}(A^h)$, up to local and
BRST-exact terms.

\section{Brief review of our unified ST--RGZ gauge fixing}
\label{sec:review}

For completeness, we recall the main ingredients of the unified ST--RGZ
gauge-fixing construction developed in
Ref.~\cite{terin2025unifiedviewpointgribovzwanzigerserreautissier}. This
framework combines two complementary ways of dealing with Gribov copies in the
Landau gauge: the copy-averaged Serreau--Tissier construction and the
Gribov--Zwanziger restriction to the first Gribov region.
In the ST formulation, one averages over the extrema $U_i$ of the Landau
functional
\begin{equation}
f[A,U]=\int d^dx\,{\rm tr}\big[(A_\mu^U)^2\big]
\end{equation}
with a non-uniform weight
\begin{equation}
\mathcal P[A,U_i]
=
\frac{\det(F[A,U_i]+\zeta\mathbf{1})}{|\det F[A,U_i]|}
\exp[-\beta f[A,U_i]],
\end{equation}
where $F[A,U_i]$ is the Hessian of the Landau functional and $\beta,\zeta$
are gauge-fixing parameters with mass dimension two. This construction does not
eliminate copies by imposing a sharp restriction on configuration space.
Instead, it modifies their statistical weight along each gauge orbit. After the
replica trick, the copy average is represented by a local nonlinear sigma model
in superspace. In the replica-symmetric branch, the resulting gluon two-point
function has a massive FP/CF-like infrared form.
The RGZ construction follows a different logic. It restricts the functional
integral to the first Gribov region by introducing the horizon functional
\begin{equation}
H(A^h)
=
g^2\int d^dx\,d^dy\,
f^{abc}A_\mu^{h,b}(x)
\big[\mathcal M^{-1}(A^h)\big]^{ad}(x,y)
f^{dec}A_\mu^{h,e}(y),
\label{eq:review-H}
\end{equation}
where \(A_\mu^h\) is transverse and gauge invariant. The operator
\begin{equation}
\mathcal M^{ab}(A^h)
=
-\partial_\mu D_\mu^{ab}(A^h)
\end{equation}
is the FP operator evaluated on $A_\mu^h$. The nonlocal
functional $H(A^h)$ can be localized through Zwanziger auxiliary fields
$(\bar\varphi,\varphi,\bar\omega,\omega)$, and its refined version includes
dimension-two condensates such as $(A^h)^2$ and
$\bar\varphi\varphi-\bar\omega\omega$. This leads to an infrared-finite
decoupling-type gluon propagator.
Our unified construction of Ref.~\cite{terin2025unifiedviewpointgribovzwanzigerserreautissier}
combines the two prescriptions by assigning to each copy the weight
\begin{equation}
\exp\big[-\beta f[A,U_i]-\gamma^4H(A^{h,U_i})\big]\,
\frac{s(i)\det(F[A,U_i]+\zeta\mathbf{1})}
{|\det F[A,U_i]|},
\label{eq:review-unified-weight}
\end{equation}
with
\begin{equation}
s(i)=\mathrm{sign}\det F[A,U_i].
\end{equation}
The corresponding average of an operator \(\mathcal O[A]\) is
\begin{equation}
\Big\langle\!\Big\langle\mathcal O[A]\Big\rangle\!\Big\rangle_{\beta,\zeta,\gamma}
=
\frac{
\displaystyle
\sum_i s(i)\,\mathcal O(A^{U_i})
\frac{\det(F[A,U_i]+\zeta\\mathbf{1})}{|\det F[A,U_i]|}
\exp[-\beta f[A,U_i]-\gamma^4H(A^{h,U_i})]
}{
\displaystyle
\sum_i s(i)
\frac{\det(F[A,U_i]+\zeta\mathbf{1})}{|\det F[A,U_i]|}
\exp[-\beta f[A,U_i]-\gamma^4H(A^{h,U_i})]
}.
\label{eq:review-copy-average}
\end{equation}
Gauge-invariant operators are unaffected by the copy average, while the limiting
cases
\begin{equation}
\gamma\to0,
\qquad
(\beta\to\infty,\zeta\to0)
\end{equation}
recover, respectively, the ST and GZ/RGZ sectors at the level of the
gauge-fixing weight.
To obtain a local action, the ST factor is localized by the replica trick and
the horizon functional is localized by Zwanziger fields. Schematically, the
resulting local action can be written as
\begin{eqnarray}
S_{\rm unif}
&=&
S_{\rm YM}
+
S_{\rm FP}
+
S_{\rm ST}[A;\{V_k\};\beta,\zeta]
\nonumber\\
&&
-\int d^dx\,
\Big[
\bar\varphi_\mu^{ac}\mathcal M^{ab}(A^h)\varphi_\mu^{bc}
-
\bar\omega_\mu^{ac}\mathcal M^{ab}(A^h)\omega_\mu^{bc}
\Big]
\nonumber\\
&&
-\gamma^2 g f^{abc}
\int d^dx\,
(A^h)_\mu^a
(\varphi+\bar\varphi)_\mu^{bc}
+
S_{\rm cond}
+
S_{\tau}
+
\cdots ,
\label{eq:review-local-action}
\end{eqnarray}
where $S_{\rm cond}$ denotes the refinement condensate sector and $S_\tau$
contains the Lagrange multiplier enforcing $\partial_\mu A_\mu^h=0$. The
ellipsis stands for the remaining replica, ghost, antighost and multiplier
terms required by locality and BRST symmetry.
In the $A^h$-based formulation, this action possesses a nilpotent
nonperturbative BRST symmetry. This is the structural reason why the construction
is algebraically controllable. The Slavnov--Taylor identity, together with the
ghost and antighost equations, constrains the counterterm and implies that the
renormalization constants of the two sectors are not independent. In particular,
the Taylor relation
\begin{equation}
Z_g Z_c Z_A^{1/2}=1
\end{equation}
is preserved, and the fields and parameters of the ST and RGZ sectors share the
same basic Yang--Mills renormalization factors.
It is important to stress that this unification is algebraic and structural. It
shows that copy averaging and horizon suppression can be embedded in a common
local BRST-invariant action. It does not, by itself, prove that the RGZ horizon
functional is dynamically generated from the ST sector, nor does it imply that
the ST and RGZ prescriptions lead to identical correlation functions. The
present work addresses a narrower question: whether the ST replica determinant
can induce, in a suitable replica branch, the leading nonlocal bilinear kernel
which has the same structure as the quadratic part of the RGZ horizon
functional.
Our unified framework contains several infrared parameters. In the RGZ sector,
the horizon parameter and refinement masses are fixed by gap equations,
\begin{equation}
\frac{\partial\Gamma_{\rm vac}}{\partial\gamma^2}=0,
\qquad
\frac{\partial\Gamma_{\rm vac}}{\partial m^2}=0,
\qquad
\frac{\partial\Gamma_{\rm vac}}{\partial M^2}=0.
\label{eq:review-RGZ-gaps}
\end{equation}
In the replica sector, the ST gap equation determines the replica curvature
$\hat\chi$. In the notation used in the ST literature, one obtains a relation
of the form
\begin{equation}
\frac{8\pi^2}{\bar g_R^2}
\frac{\beta_R}{\bar\mu^2}
=
(\hat\chi_R+\zeta_R)
\ln\frac{\hat\chi_R+\zeta_R}{\bar\mu^2}
-
\hat\chi_R
\ln\frac{\hat\chi_R}{\bar\mu^2},
\label{eq:review-ST-gap}
\end{equation}
with $\hat\chi_R\ge0$ in the replica-symmetric branch.
The two relevant infrared realizations are then distinguished by the behavior of
$\hat\chi_R$. For $\hat\chi_R>0$, the replica determinant produces a local
screening contribution and the ST sector reduces to a CF-like massive
description. In the limiting branch $\hat\chi_R=0$, the local replica
curvature vanishes and the determinant expansion develops nonlocal contributions
built from $\mathcal M^{-1}(A^h)$. It is this second branch that is analyzed in
the following sections.
The effective transverse gluon propagator in the unified framework can be
parametrized as
\begin{equation}
D_T(p^2)
=
\frac{p^2+M^2}
{(p^2+M^2)(p^2+\mu_{\rm IR}^2)+\lambda^4},
\label{eq:review-DT}
\end{equation}
where
\begin{equation}
\mu_{\rm IR}^2
=
m^2+\Xi_{\rm rep}\,\beta,
\label{eq:review-muIR}
\end{equation}
and
\begin{equation}
\lambda^4
=
2g^2N\,\gamma_{\rm eff}^4.
\label{eq:review-lambda}
\end{equation}
The phase selector \(\Xi_{\rm rep}\) is defined by
\begin{equation}
\Xi_{\rm rep}
=
\begin{cases}
1, & \text{replica-symmetric branch }(\hat\chi_R>0),\\[4pt]
0, & \text{replica-broken branch }(\hat\chi_R=0).
\end{cases}
\label{eq:review-Xi}
\end{equation}
Thus, in the symmetric branch, the ST sector contributes through a local
screening mass. In the broken branch, the local ST screening contribution is
absent and the relevant induced effect is encoded instead in the horizon-like
scale contained in $\lambda^4$.
This notation makes explicit how double counting is avoided. The parameter
$\beta$ and the induced horizon contribution are not treated as two
independent ST-generated infrared effects simultaneously active in the same
replica phase. The local CF-like mass appears for $\Xi_{\rm rep}=1$, whereas
the induced horizon-like kernel belongs to the branch with $\Xi_{\rm rep}=0$.
If an explicit RGZ horizon term is included together with the induced
contribution generated by the replica sector, the effective horizon strength is
written as
\begin{equation}
\gamma_{\rm eff}^4
=
\gamma^4+\gamma_{\rm ind}^4.
\label{eq:review-gamma-eff}
\end{equation}
The standard RGZ horizon condition then applies to $\gamma_{\rm eff}$, while
$\gamma_{\rm ind}$ is fixed by the ST determinant once the replica branch and
the regulator $\zeta$ are specified.
The rest of the paper focuses on the derivation and interpretation of
$\gamma_{\rm ind}$. More precisely, we show that the small-$\zeta$ expansion
of the replica determinant induces a nonlocal bilinear term proportional to the
quadratic horizon kernel $H_2(A^h)$. This establishes an effective
leading-order link between the ST replica-broken sector and the RGZ infrared
kernel, without claiming a reconstruction of the full nonlinear functional
$H(A^h)$.

\section{Effective replica action and determinant expansion}
\label{sec:replica-eff}

We now analyze the contribution generated by the ST replica sector after the
nonlinear sigma model superfields are integrated out. The purpose of this
section is to identify the part of the resulting determinant that can generate
a nonlocal gluonic kernel and to clarify the sense in which this kernel matches
the quadratic part of the Gribov horizon functional.
The ST replica construction admits different infrared realizations. In the
replica-symmetric branch, characterized by a positive replica curvature
$\hat\chi_R>0$, the determinant generated by the nonlinear sigma model fields
produces a local screening contribution. In this branch the $n\to0$ limit
reduces, for the gluon and FP ghost sector, to a massive FP/CF-type
description. In contrast, in the limiting branch $\hat\chi_R=0$, the local
replica curvature vanishes and the determinant expansion develops nonlocal
contributions involving inverse powers of the FP operator. This is the branch
relevant for the induced horizon-like kernel considered below.
Before implementing the replica trick, the ST gauge fixing can be represented
schematically as
\begin{equation}
S_{\mathrm{ST}}[A,c,\bar c,b;U]
=
\int d^dx\,
\left[
\bar c^a\,\partial_\mu D_\mu^{ab}(A)\,c^b
+\zeta\,\bar c^a c^a
+i\,b^a\,\partial_\mu A_\mu^a
+\frac{\beta}{2}\,A_\mu^a A_\mu^a
\right],
\label{eq:ST-action-before-replica}
\end{equation}
where $U(x)\in SU(N)$ labels copies along the gauge orbit and
$\beta,\zeta$ are ST gauge-fixing parameters. After the replica trick and the
integration over the replica superfields, the replica sector gives an effective
contribution depending on the gauge background. In the BRST-invariant
formulation, this dependence is expressed in terms of the transverse
gauge-invariant field $A_\mu^h$,
\begin{equation}
S_{\mathrm{rep}}^{\mathrm{eff}}[A^h]
=
\frac{\beta}{2}\int d^dx\,(A_\mu^h)^2
-
\left(
\ln\det[\mathcal M(A^h)+\zeta\,\mathbf{1}] 
-
\ln|\det\mathcal M(A^h)|
\right),
\label{eq:Srep-eff}
\end{equation}
where
\begin{equation}
\mathcal M^{ab}(A^h)
=
-\partial_\mu D_\mu^{ab}(A^h)
\end{equation}
is the FP operator evaluated on $A_\mu^h$.
To isolate the nonlocal part induced by the regulator $\zeta$, we use
\begin{equation}
\ln\det[\mathcal M+\zeta]
=
\ln\det\mathcal M
+
\mathrm{Tr}\ln(1+\zeta\,\mathcal M^{-1}).
\end{equation}
For small \(\zeta\), this gives
\begin{equation}
\ln\det[\mathcal M+\zeta]-\ln\det\mathcal M
=
\zeta\,\mathrm{Tr}\,\mathcal M^{-1}
-
\frac{\zeta^2}{2}\,\mathrm{Tr}\,\mathcal M^{-2}
+
O(\zeta^3).
\label{eq:det-expansion-zeta}
\end{equation}
Therefore,
\begin{equation}
S_{\mathrm{rep}}^{\mathrm{eff}}[A^h]
=
\frac{\beta}{2}\int d^dx\,(A_\mu^h)^2
-
\zeta\,\mathrm{Tr}\,\mathcal M^{-1}(A^h)
+
\frac{\zeta^2}{2}\,\mathrm{Tr}\,\mathcal M^{-2}(A^h)
+
O(\zeta^3).
\label{eq:Srep-zeta-expanded}
\end{equation}
The first term in Eq.~\eqref{eq:Srep-zeta-expanded} is local and is responsible
for the CF-like screening contribution in the replica-symmetric branch. The
nonlocal horizon-like contribution arises from the determinant terms involving
inverse powers of $\mathcal M(A^h)$. In the present work we keep the leading
term linear in $\zeta$, since it is the first term capable of generating the
nonlocal bilinear kernel relevant for comparison with the RGZ horizon sector.
The higher-order terms in $\zeta$, starting with
$\frac{\zeta^2}{2}\mathrm{Tr}\,\mathcal M^{-2}$, generate subleading nonlocal
kernels and local counterterms. They may renormalize the matching coefficient
defined below, but they do not change the leading linear-$\zeta$ identification
of the bilinear kernel.
We expand the FP operator around its free part,
\begin{equation}
\mathcal M(A^h)
=
\mathcal M_0+V[A^h],
\qquad
\mathcal M_0^{ab}
=
-\partial^2\delta^{ab},
\qquad
V^{ab}[A^h]
=
-g f^{acb}A_\mu^{h,c}\partial_\mu .
\label{eq:M0V}
\end{equation}
The inverse operator is then
\begin{equation}
\mathcal M^{-1}
=
\mathcal M_0^{-1}
-
\mathcal M_0^{-1}V\,\mathcal M_0^{-1}
+
\mathcal M_0^{-1}V\,\mathcal M_0^{-1}V\,\mathcal M_0^{-1}
+\cdots .
\label{eq:M-inverse-expansion}
\end{equation}
The term independent of $A^h$ contributes only to the vacuum functional, the
term linear in $A^h$ vanishes after the color trace because it contains a
single structure constant, the leading nontrivial gluonic contribution is
therefore quadratic in $A^h$. Keeping this term, the part of the effective
action linear in $\zeta$ gives
\begin{equation}
\left[
-\zeta\,\mathrm{Tr}\,\mathcal M^{-1}(A^h)
\right]_{A^2}
=
-\zeta\,
\mathrm{Tr}
\left(
\mathcal M_0^{-1}V[A^h]\mathcal M_0^{-1}
V[A^h]\mathcal M_0^{-1}
\right).
\label{eq:bilinear-trace}
\end{equation}
In coordinate space this contribution can be written as
\begin{eqnarray}
\left[
-\zeta\,\mathrm{Tr}\,\mathcal M^{-1}(A^h)
\right]_{A^2}
&=&
\zeta g^2
\int d^dx\,d^dy\,
A_\mu^{h,a}(x)\,
\mathcal K_{\mu\nu}^{ab}(x,y)\,
A_\nu^{h,b}(y),
\label{eq:K-coordinate-form}
\end{eqnarray}
with
\begin{eqnarray}
\mathcal K_{\mu\nu}^{ab}(x,y)
&=&
f^{acm}f^{bdm}
\left[
\partial_\mu^x\mathcal M_0^{-1}(x-y)
\right]
\mathcal M_0^{-1}(y-x)
\left[
\partial_\nu^y\mathcal M_0^{-1}(y-x)
\right],
\label{eq:K-coordinate}
\end{eqnarray}
up to local terms and normalization conventions. Equivalently, in momentum
space,
\begin{equation}
\left[
-\zeta\,\mathrm{Tr}\,\mathcal M^{-1}(A^h)
\right]_{A^2}
=
\zeta g^2 C_A
\int\frac{d^dp}{(2\pi)^d}\,
A_\mu^{h,a}(p)\,
\mathcal K_{\mu\nu}(p)\,
A_\nu^{h,a}(-p),
\label{eq:K-momentum-form}
\end{equation}
where \(C_A=N\) for \(SU(N)\). The kernel can be decomposed as
\begin{equation}
\mathcal K_{\mu\nu}(p)
=
P_{\mu\nu}(p)\,\mathcal K_T(p^2)
+
\text{local and longitudinal terms},
\qquad
P_{\mu\nu}(p)
=
\delta_{\mu\nu}
-
\frac{p_\mu p_\nu}{p^2}.
\label{eq:K-decomposition}
\end{equation}
Since $A_\mu^h$ is transverse, the longitudinal part does not contribute to
the infrared transverse two-point function.
The quadratic part of the BRST-invariant Gribov horizon functional is obtained
by replacing the full inverse operator in $H(A^h)$ by the free inverse
operator $\mathcal M_0^{-1}$. Thus,
\begin{equation}
H(A^h)
=
H_2(A^h)
+
O((A^h)^3),
\end{equation}
with
\begin{equation}
H_2(A^h)
=
g^2 C_A
\int\frac{d^dp}{(2\pi)^d}\,
A_\mu^{h,a}(p)\,
\frac{P_{\mu\nu}(p)}{p^2}\,
A_\nu^{h,a}(-p),
\label{eq:H2-momentum-sec3}
\end{equation}
again up to conventional normalizations of the generators and of the FP
operator.
Equations~\eqref{eq:K-momentum-form}--\eqref{eq:H2-momentum-sec3} define the
precise sense in which the ST determinant induces an RGZ-type horizon kernel.
The correspondence is obtained by projecting the nonlocal transverse part of
$\mathcal K_{\mu\nu}(p)$ onto the structure $P_{\mu\nu}(p)/p^2$. This
projection defines the induced Gribov scale. In terms of renormalized
quantities,
\begin{equation}
\gamma_{\mathrm{ind},R}^{4}
=
\kappa_R(d,N,\mu)\,\mu^2\,\zeta_R
+
O(\zeta_R^2).
\label{eq:gamma-ind-kappa-sec3}
\end{equation}
The scale $\mu$ is included to make the dimensions explicit when
$\zeta_R$ has mass dimension two. The coefficient
$\kappa_R(d,N,\mu)$ is not universal: it depends on the subtraction scheme,
the normalization of $H_2(A^h)$, and the precise projection used to isolate
the nonlocal transverse kernel. Its physical sign is fixed by requiring that
the induced term enter the quadratic action with the same sign as the usual
RGZ horizon contribution, i.e. $\gamma_{\mathrm{ind},R}^{4}>0$ in the
horizon-dominated branch.
This result should not be interpreted as a reconstruction of the full nonlinear
functional $H(A^h)$. The full horizon functional contains the complete
operator $\mathcal M^{-1}(A^h)$, and therefore includes infinitely many
higher-order vertices in $A^h$. The present derivation establishes an
effective leading-order correspondence between the ST determinant and the RGZ
horizon structure at the level of the nonlocal bilinear kernel. This is the
part relevant for the tree-level infrared propagator and for defining the
induced scale $\gamma_{\mathrm{ind}}$.
It is also useful to emphasize that the induced contribution is controlled by
the replica branch. In the replica-symmetric phase, the determinant is dominated
by the local curvature $\hat\chi_R>0$, and the ST sector yields a local
screening mass. In the replica-broken branch, $\hat\chi_R=0$, the nonlocal
part of the determinant becomes the relevant infrared contribution. The
resulting effective action may then be summarized, at leading order, as
\begin{equation}
S_{\mathrm{rep}}^{\mathrm{eff}}[A^h]
=
\frac{\beta}{2}\int d^dx\,(A_\mu^h)^2
+
\gamma_{\mathrm{ind}}^4\,H_2(A^h)
+
\text{local terms}
+
O((A^h)^3,\zeta^2).
\label{eq:Srep-summary}
\end{equation}
In the next section we discuss the localization of this induced quadratic
horizon kernel and its relation to the standard RGZ form.

\section{Emergence of the horizon kernel}
\label{sec:horizon-functional}

The determinant expansion derived in Sec.~\ref{sec:replica-eff} shows that the
ST replica sector generates, in the replica-broken branch, a nonlocal bilinear
kernel built from inverse FP operators. We now clarify how this kernel is
related to the BRST-invariant Gribov horizon structure and how it can be
localized in an RGZ-like form.
The full BRST-invariant horizon functional is
\begin{equation}
H(A^h)
=
g^2\int d^dx\,d^dy\,
f^{abc}A_\mu^{h,b}(x)
\big[\mathcal M^{-1}(A^h)\big]^{ad}(x,y)
f^{dec}A_\mu^{h,e}(y).
\label{eq:horizon-full-sec4}
\end{equation}
It is a nonlinear functional of $A^h$, because the inverse FP operator itself
depends on $A^h$. Expanding it around the free FP operator gives
\begin{equation}
H(A^h)
=
H_2(A^h)
+
H_3(A^h)
+
H_4(A^h)
+\cdots ,
\label{eq:horizon-expansion-sec4}
\end{equation}
where $H_n(A^h)$ denotes the contribution of order $n$ in $A^h$. The
quadratic contribution is
\begin{equation}
H_2(A^h)
=
g^2 C_A
\int\frac{d^dp}{(2\pi)^d}\,
A_\mu^{h,a}(p)
\frac{P_{\mu\nu}(p)}{p^2}
A_\nu^{h,a}(-p),
\label{eq:H2-sec4}
\end{equation}
up to normalization conventions.
The result of Sec.~\ref{sec:replica-eff} is that the leading nonlocal part of
the ST determinant in the replica-broken branch reproduces the same bilinear
color and Lorentz structure as Eq.~\eqref{eq:H2-sec4}. Thus, what emerges from
the ST sector at the order considered here is not the full nonlinear functional
$H(A^h)$, but its nonlocal quadratic kernel $H_2(A^h)$. This is the part
which controls the tree-level infrared modification of the transverse gluon
propagator.
More explicitly, the induced contribution can be written in the form
\begin{equation}
S_{\rm ind}^{(2)}[A^h]
=
\gamma_{\mathrm{ind}}^4\,H_2(A^h),
\label{eq:Sind-H2}
\end{equation}
with
\begin{equation}
\gamma_{\mathrm{ind},R}^4
=
\kappa_R(d,N,\mu)\,\mu^2\,\zeta_R
+
O(\zeta_R^2).
\label{eq:gamma-ind-sec4}
\end{equation}
Here $\kappa_R(d,N,\mu)$ is defined by the transverse projection of the induced
nonlocal kernel onto $P_{\mu\nu}(p)/p^2$, as described in
Sec.~\ref{sec:replica-eff}. It depends on the renormalization prescription and
on the normalization convention used for the horizon functional.
The sign of $\kappa_R$ is fixed by the requirement that the induced kernel
belongs to the horizon-dominated branch. In this branch the induced
contribution must enter the quadratic transverse action with the same sign as
the usual RGZ horizon term. Equivalently, one requires
$\gamma_{\mathrm{ind},R}^4>0$. This condition is not an additional gap
equation; it selects the infrared branch in which the determinant-induced
kernel has the RGZ-type sign.
The bilinear kernel \eqref{eq:Sind-H2} can be localized by introducing
Zwanziger-type auxiliary fields
$(\bar\varphi,\varphi,\bar\omega,\omega)$. At quadratic order, the
localization is of the standard form
\begin{eqnarray}
\gamma_{\mathrm{ind}}^4 H_2(A^h)
&\longrightarrow&
\int d^dx\,
\Big[
\bar\varphi_\mu^{ac}\,\mathcal M_0^{ab}\,\varphi_\mu^{bc}
-
\bar\omega_\mu^{ac}\,\mathcal M_0^{ab}\,\omega_\mu^{bc}
\Big]
\nonumber\\
&&
-
g\,\gamma_{\mathrm{ind}}^2 f^{abc}
\int d^dx\,
A_\mu^{h,a}
(\varphi+\bar\varphi)_\mu^{bc}.
\label{eq:localization-H2}
\end{eqnarray}
If one promotes $\mathcal M_0$ to the full operator $\mathcal M(A^h)$, this
takes the standard RGZ local form. In the present derivation, however, this
promotion should be understood as the usual RGZ completion of the bilinear
kernel, not as something derived from the ST determinant to all orders in
$A^h$.
This distinction is important. The ST determinant calculation establishes a
controlled leading-order correspondence:
\begin{equation}
-\zeta\,\mathrm{Tr}\,\mathcal M^{-1}(A^h)
\quad
\Longrightarrow
\quad
\gamma_{\mathrm{ind}}^4\,H_2(A^h)
+
\text{local terms}
+
O((A^h)^3,\zeta^2).
\label{eq:effective-correspondence}
\end{equation}
The RGZ local action can then be used as an effective infrared completion of
this induced kernel. The present work therefore provides a microscopic origin
for the leading horizon-like bilinear structure, rather than a full derivation
of the complete RGZ horizon functional.
If an explicit RGZ horizon term is already present in the starting action, the
induced contribution simply shifts the effective horizon strength:
\begin{equation}
\gamma_{\mathrm{eff}}^4
=
\gamma^4+\gamma_{\mathrm{ind}}^4.
\label{eq:gamma-eff-sec4}
\end{equation}
The ordinary RGZ horizon condition then applies to $\gamma_{\mathrm{eff}}$.
By contrast, $\gamma_{\mathrm{ind}}$ is fixed by the ST sector through the
determinant expansion and the replica branch. Therefore, the induced scale is
not determined by a new independent horizon condition; it is inherited from the
ST regulator $\zeta$ and from the projection coefficient $\kappa_R$.
At leading order in the small-$\zeta$ expansion, the relation
$\gamma_{\mathrm{ind}}^4\propto\zeta$ is structurally stable under
renormalization. Indeed, writing
\begin{equation}
\zeta_0=Z_\zeta\,\zeta_R,
\qquad
\gamma_{\mathrm{ind},0}^4
=
Z_{\gamma_{\mathrm{ind}}}\,
\gamma_{\mathrm{ind},R}^4,
\end{equation}
one obtains
\begin{equation}
\gamma_{\mathrm{ind},R}^4
=
Z_{\gamma_{\mathrm{ind}}}^{-1}Z_\zeta\,
\kappa_0(d,N,\mu)\,\mu^2\zeta_R
+
O(\zeta_R^2)
\equiv
\kappa_R(d,N,\mu)\,\mu^2\zeta_R
+
O(\zeta_R^2).
\label{eq:kappa-renormalization-sec4}
\end{equation}
Thus renormalization changes the coefficient $\kappa_R$, but not the
existence of the leading linear relation between $\gamma_{\mathrm{ind}}^4$
and $\zeta$. Beyond leading order, higher powers of $\zeta$ generate
additional nonlocal structures and finite scheme-dependent corrections to the
matching.
The physical interpretation is therefore as follows. In the replica-symmetric
branch, the ST determinant produces a local screening mass and no induced
horizon kernel. In the replica-broken branch, the same determinant produces a
nonlocal bilinear kernel with the RGZ horizon structure. The ST parameters
$(\beta,\zeta)$ therefore control two different infrared mechanisms in two
different replica sectors:
\begin{equation}
\hat\chi_R>0:
\quad
\text{local CF-like screening},
\qquad
\hat\chi_R=0:
\quad
\text{induced RGZ-type horizon kernel}.
\end{equation}
This phase-dependent interpretation is summarized by the chain
\begin{equation}
\text{ST replica determinant}
\quad
\xrightarrow[\hat\chi_R=0]{\text{small }\zeta}
\quad
\text{nonlocal bilinear kernel}
\quad
\xrightarrow{\text{projection}}
\quad
\gamma_{\mathrm{ind}}^4H_2(A^h).
\end{equation}
It is in this restricted and precise sense that an RGZ-type horizon structure
emerges from the ST replica sector.


\section{Induced horizon coupling: limits, interpretation, and relation to RGZ}
\label{sec:limits-interpretation}

The previous sections identified the leading nonlocal bilinear kernel induced by
the ST determinant in the replica-broken branch. We now clarify the limiting
cases, the interpretation of the induced scale, and the relation between this
scale and the standard RGZ Gribov parameter.
The induced contribution is phase dependent. In the replica-symmetric branch,
$\hat\chi_R>0$, the ST determinant is dominated by the local curvature of the
replica sector. The resulting infrared effect is a local CF-like screening mass.
In this case, the determinant does not generate the nonlocal horizon kernel
discussed above. In the replica-broken branch, represented by the limiting
condition $\hat\chi_R=0$, the local curvature vanishes and the determinant
develops nonlocal contributions involving $\mathcal M^{-1}(A^h)$. At leading
order in $\zeta$ and at quadratic order in $A^h$, this gives
\begin{equation}
\Gamma_{\rm eff}[A^h]
\supset
\gamma_{\mathrm{ind}}^4\,H_2(A^h),
\label{eq:limits-induced-H2}
\end{equation}
where
\begin{equation}
H_2(A^h)
=
g^2 C_A
\int\frac{d^dp}{(2\pi)^d}
A_\mu^{h,a}(p)
\frac{P_{\mu\nu}(p)}{p^2}
A_\nu^{h,a}(-p).
\label{eq:limits-H2}
\end{equation}
Thus, the relation between the ST and RGZ sectors is not a statement of full
dynamical equivalence. It is an effective correspondence between the leading
nonlocal bilinear kernel generated by the ST determinant and the quadratic
kernel of the BRST-invariant RGZ horizon functional. The higher nonlinear
terms of $H(A^h)$ are not derived here from the ST determinant to all orders.
They belong to the standard RGZ completion of the horizon sector.
The induced scale is parametrized by
\begin{equation}
\gamma_{\mathrm{ind},R}^{4}
=
\kappa_R(d,N,\mu)\,\mu^2\zeta_R
+
O(\zeta_R^2).
\label{eq:limits-gamma-ind}
\end{equation}
The coefficient $\kappa_R(d,N,\mu)$ is fixed by matching the transverse
nonlocal part of the induced kernel to the structure $P_{\mu\nu}(p)/p^2$.
It is therefore a projection-dependent and scheme-dependent coefficient, not a
universal number. Its expected sign in the horizon-dominated branch is positive,
in the sense that the induced term contributes with the same sign as the usual
RGZ horizon term in the quadratic transverse action.
There is no additional independent horizon condition for
$\gamma_{\mathrm{ind}}$. In the standard RGZ framework, the Gribov parameter
is fixed by the horizon condition. In the present mechanism,
$\gamma_{\mathrm{ind}}$ is instead fixed by the ST sector: once the replica
branch, the regulator $\zeta_R$, and the renormalization prescription are
chosen, the induced contribution is determined by the matching coefficient
$\kappa_R$. If an explicit RGZ horizon term is also included, then the
standard RGZ gap equation applies to the effective combination
\begin{equation}
\gamma_{\mathrm{eff}}^4
=
\gamma^4+\gamma_{\mathrm{ind}}^4.
\label{eq:limits-gamma-eff}
\end{equation}
In this case, the induced scale acts as a microscopic contribution to the
effective horizon strength, while the usual RGZ horizon condition still fixes
the full infrared configuration.
The topological ST limit is recovered when the regulator is removed:
\begin{equation}
\zeta_R\to0
\qquad\Longrightarrow\qquad
\gamma_{\mathrm{ind},R}^4\to0.
\end{equation}
Therefore, no horizon-like kernel is induced in the strict $\zeta_R=0$ limit. For
finite $\zeta_R$, the induced term is present only in the replica-broken
branch. This makes clear that $\zeta_R$ controls the strength of the induced
nonlocality, while $\beta_R$ controls the local copy-weighting scale.
The phase dependence is crucial for avoiding double counting of infrared
effects. The ST sector does not simultaneously generate both a local CF mass
and an induced horizon kernel within the same infrared realization. Instead,
one has
\begin{equation}
\begin{array}{lll}
\hat\chi_R>0:
& \text{replica-symmetric branch}
& \Longrightarrow\quad
\text{local CF-like screening},\\[4pt]
\hat\chi_R=0:
& \text{replica-broken branch}
& \Longrightarrow\quad
\text{induced RGZ-type horizon kernel}.
\end{array}
\label{eq:phase-branches}
\end{equation}
This separation is implemented in the effective propagator by introducing the
phase selector $\Xi_{\rm rep}$. The local ST mass enters the transverse
inverse propagator only when $\Xi_{\rm rep}=1$, whereas the induced horizon
contribution belongs to the branch with $\Xi_{\rm rep}=0$.
Accordingly, the infrared mass parameter entering the transverse propagator is
written as
\begin{equation}
\mu_{\rm IR}^2
=
m^2+\Xi_{\rm rep}\,\beta_R,
\label{eq:limits-muIR}
\end{equation}
with
\begin{equation}
\Xi_{\rm rep}
=
\begin{cases}
1, & \hat\chi_R>0,\\[4pt]
0, & \hat\chi_R=0.
\end{cases}
\label{eq:limits-Xi}
\end{equation}
The horizon strength is encoded separately in
\begin{equation}
\lambda^4
=
2g^2N\,\gamma_{\rm eff}^4,
\qquad
\gamma_{\rm eff}^4
=
\gamma^4+\gamma_{\rm ind}^4.
\label{eq:limits-lambda}
\end{equation}
Thus, the local CF-like scale and the horizon-like scale are not counted twice.
They correspond to distinct infrared mechanisms selected by the replica branch.
Under renormalization, the leading proportionality
$\gamma_{\mathrm{ind}}^4\propto\zeta$ is expected to remain stable in the
sense that the same operator mixing pattern is preserved at leading order in
the small-$\zeta$ expansion. Renormalization changes the coefficient
$\kappa_R(d,N,\mu)$, and higher powers of $\zeta_R$ generate additional
subleading nonlocal structures. Therefore,
\begin{equation}
\gamma_{\mathrm{ind},R}^{4}
=
\kappa_R(d,N,\mu)\,\mu^2\zeta_R
+
\kappa_R^{(2)}(d,N,\mu)\,\zeta_R^2
+
\cdots .
\label{eq:limits-higher-zeta}
\end{equation}
The present work keeps only the first term in this expansion. A full
renormalization-group analysis of the tower of $\zeta_R$-dependent nonlocal
operators is beyond the scope of this paper, but the leading correspondence is
not spoiled by finite renormalizations; it is simply absorbed into the
definition of $\kappa_R$.
The interpretation of the induced scale is therefore precise. It is not a new
fundamental RGZ parameter introduced by hand. It is the coefficient of the
leading nonlocal bilinear horizon kernel generated by the ST determinant in the
replica-broken branch. Conversely, the usual RGZ parameter $\gamma$, when
present, retains its standard meaning as the parameter constrained by the
horizon condition. The two contributions combine only at the level of the
effective horizon strength.
This distinction also clarifies the role of our unified ST--RGZ framework. Our
unified action provides a local BRST-invariant environment in which the ST and
RGZ sectors can be discussed consistently. The present mechanism adds a
dynamical interpretation: part of the horizon-like infrared structure can be
understood as an induced kernel generated by the ST replica sector itself.
However, this interpretation is limited to the leading nonlocal bilinear
structure and should not be read as a complete derivation of the full nonlinear
RGZ action from the ST construction.


\section{Tree-level gluon propagator and phase-dependent infrared structure}
\label{sec:gluon-propagator}

We now derive the tree-level transverse gluon propagator associated with the
effective infrared structure discussed above. The purpose of this section is
not to provide a full phenomenological fit, but to show how the phase dependence
of the replica sector is reflected in the infrared form of the two-point
function and how double counting of infrared scales is avoided.
In a linear covariant gauge, the gluon two-point function can be decomposed as
\begin{equation}
D_{\mu\nu}^{ab}(p)
=
\delta^{ab}
\left[
P_{\mu\nu}(p)\,D_T(p^2)
+
\alpha\,\frac{p_\mu p_\nu}{p^4}
\right],
\qquad
P_{\mu\nu}(p)
=
\delta_{\mu\nu}
-
\frac{p_\mu p_\nu}{p^2}.
\label{eq:prop-decomposition}
\end{equation}
Since the induced horizon-like term is written in terms of the transverse
gauge-invariant field $A_\mu^h$, the nontrivial infrared information is
contained in the transverse form factor $D_T(p^2)$.
At the level of the quadratic effective action, the transverse inverse
propagator takes the form
\begin{equation}
\Gamma_T(p^2)
=
\frac{(p^2+M^2)(p^2+\mu_{\rm IR}^2)+\lambda^4}
{p^2+M^2},
\label{eq:GammaT}
\end{equation}
so that
\begin{equation}
D_T(p^2)
=
\frac{p^2+M^2}
{(p^2+M^2)(p^2+\mu_{\rm IR}^2)+\lambda^4}.
\label{eq:DT-general}
\end{equation}
Here $M^2$ denotes the usual RGZ refinement scale associated with the
Zwanziger auxiliary sector, while
\begin{equation}
\mu_{\rm IR}^2
=
m^2+\Xi_{\rm rep}\,\beta_R
\label{eq:muIR-sec6}
\end{equation}
contains the local infrared mass contributions. The phase selector
$\Xi_{\rm rep}$ is defined as
\begin{equation}
\Xi_{\rm rep}
=
\begin{cases}
1, & \text{replica-symmetric branch }(\hat\chi_R>0),\\[4pt]
0, & \text{replica-broken branch }(\hat\chi_R=0).
\end{cases}
\label{eq:Xi-sec6}
\end{equation}
The horizon strength is encoded in
\begin{equation}
\lambda^4
=
2g^2N\,\gamma_{\rm eff}^4,
\qquad
\gamma_{\rm eff}^4
=
\gamma^4+\gamma_{\rm ind}^4.
\label{eq:lambda-sec6}
\end{equation}
If no explicit RGZ horizon term is introduced, one simply sets
$\gamma=0$, so that $\gamma_{\rm eff}=\gamma_{\rm ind}$.
Equations~\eqref{eq:muIR-sec6}--\eqref{eq:lambda-sec6} implement the central
phase-dependent structure of the paper. In the replica-symmetric branch,
$\Xi_{\rm rep}=1$, the ST sector contributes through the local screening mass
$\beta_R$. In the replica-broken branch, $\Xi_{\rm rep}=0$, this local
screening contribution is absent, and the ST sector contributes instead
through the induced horizon-like scale $\gamma_{\rm ind}$. Thus the same
replica sector is not counted twice in the transverse propagator.
The two limiting forms are immediate. In the replica-symmetric branch, with no
horizon term,
\begin{equation}
\lambda^4\to0,
\qquad
\Xi_{\rm rep}=1,
\qquad
m^2\to0,
\end{equation}
one obtains the CF/massive FP form
\begin{equation}
D_T^{\rm ST}(p^2)
=
\frac{1}{p^2+\beta_R}.
\label{eq:DT-ST}
\end{equation}
This is the standard infrared realization of the ST construction in the
replica-symmetric phase.
In the horizon-dominated branch, the local ST screening contribution is absent,
\[
\Xi_{\rm rep}=0,
\]
and the propagator becomes
\begin{equation}
D_T^{\rm RGZ-like}(p^2)
=
\frac{p^2+M^2}
{(p^2+M^2)(p^2+m^2)+\lambda^4}.
\label{eq:DT-RGZ-like}
\end{equation}
This is the usual RGZ decoupling form, with the important distinction that
$\lambda^4$ may contain an induced part generated by the ST determinant,
\[
\lambda^4
=
2g^2N(\gamma^4+\gamma_{\rm ind}^4).
\]
Therefore, the induced mechanism does not add an independent mass term on top
of the CF mass in the same phase. Instead, it supplies a horizon-like
contribution in the replica-broken branch. This is the practical sense in which
the effective propagator avoids double counting of infrared effects.
At zero momentum, the general transverse propagator gives
\begin{equation}
D_T(0)
=
\frac{M^2}
{M^2\mu_{\rm IR}^2+\lambda^4}.
\label{eq:DT-zero}
\end{equation}
Thus the propagator is infrared finite as long as
\[
M^2\mu_{\rm IR}^2+\lambda^4\neq0.
\]
In the replica-symmetric branch, finiteness is controlled by the local screening
scale $\beta_R$. In the replica-broken branch, finiteness is controlled by the
horizon-like scale $\lambda^4$, which may include the induced contribution
$\gamma_{\rm ind}^4$.
The pole structure of the horizon-dominated propagator follows from the
denominator
\begin{equation}
Q(p^2)
=
(p^2+M^2)(p^2+m^2)+\lambda^4.
\label{eq:Q-sec6}
\end{equation}
For parameters in the usual decoupling regime, this quadratic polynomial in
\(p^2\) factorizes as
\begin{equation}
Q(p^2)
=
(p^2+m_+^2)(p^2+m_-^2),
\qquad
m_-^2=(m_+^2)^*,
\label{eq:complex-poles-sec6}
\end{equation}
giving a pair of complex-conjugate poles. These poles indicate positivity
violation of the elementary gluon propagator and should not be interpreted as
physical asymptotic particles. Their role is to provide an effective infrared
input for gauge-invariant observables.
The induced scale $\gamma_{\rm ind}$ affects this pole structure through
$\lambda^4$. Therefore, possible continuum or lattice tests of the mechanism
may focus on quantities sensitive to the infrared curvature of the transverse
propagator, the zero-momentum value $D_T(0)$, and the location of complex-pole
fits in RGZ-inspired parametrizations. In a copy-weighted lattice setup, one
could in principle vary the copy-weighting parameters and test whether the
effective horizon scale changes consistently with the predicted
$\zeta_R$-dependence.
The phase-dependent propagator can be summarized as
\begin{equation}
D_T(p^2)
=
\begin{cases}
\displaystyle
\frac{1}{p^2+\beta_R},
&
\text{replica-symmetric branch, CF-like screening},
\\[14pt]
\displaystyle
\frac{p^2+M^2}
{(p^2+M^2)(p^2+m^2)+2g^2N\gamma_{\rm eff}^4},
&
\text{replica-broken branch, RGZ-type horizon kernel}.
\end{cases}
\label{eq:phase-propagator-summary}
\end{equation}
This expression is not meant to imply that the two branches are simultaneously
realized. Rather, it displays the two possible infrared outputs of the ST
replica sector. The local CF-like mass and the induced horizon-like kernel are
alternative phase-dependent realizations of the same copy-averaging dynamics.
This structure admits a natural statistical interpretation.  
The ST parameter $\beta$ controls the distribution of gauge copies along a
gauge orbit in a way analogous to a parameter controlling the width of a
statistical ensemble, while the horizon scale controls the suppression of
configurations near the boundary of the Gribov region. For small infrared
control scales, many copies contribute with comparable weight, producing a
broad copy ensemble. When the replica curvature vanishes and the system enters
the replica-broken branch, the determinant develops a nonlocal component and an
RGZ-type horizon kernel is induced. Table~\ref{tab:IsingUnified} summarizes
this correspondence.
\begin{table*}
\centering
\small
\caption{(a) Statistical analogy between the Ising model and the unified
ST--RGZ framework; (b) phase structure and order parameters in the Ising and
ST--RGZ systems. The analogy is only qualitative: the CF-like screening mass
and the RGZ-type horizon kernel correspond to distinct replica branches and
are not simultaneous independent infrared contributions.}
\label{tab:IsingUnified}

\textbf{(a) Statistical analogy}\\[4pt]

\begin{tabular}{l p{4cm} p{6cm}}
\toprule
\textbf{Feature} & \textbf{Ising model} & \textbf{Unified ST--RGZ framework} \\ 
\midrule

Microscopic states &
Spins \(\sigma_i=\pm1\) &
Gauge copies \(A_\mu^U\) along a gauge orbit \\[4pt]

Energy functional &
\(H(\sigma)\) &
\(\mathcal{E}[A^U] = \beta\,\mathcal{F}[A,U] + \gamma^4 H(A^h)\) \\[4pt]

Statistical weight &
\(e^{-\beta H(\sigma)}\) &
\(e^{-\beta\mathcal{F}[A,U] - \gamma^4 H(A^h)}\) \\[4pt]

Control parameters &
\(\beta^{-1}\) (temperature) &
\(\beta^{-1}\) (copy smoothing), \(\gamma\) (horizon strength) \\[4pt]

Partition function &
\(Z = \sum_\sigma e^{-\beta H}\) &
\(Z = \int[dA\, dU]\, e^{-\beta\mathcal{F}[A,U] - \gamma^4 H(A^h)}\) \\[4pt]

\bottomrule
\end{tabular}

\vspace{0.7cm}

\textbf{(b) Phase structure and order parameters}\\[4pt]

\begin{tabular}{l p{4cm} p{6cm}}
\toprule
\textbf{Feature} & \textbf{Ising model} & \textbf{Unified ST--RGZ framework} \\
\midrule

Replica-symmetric branch &
Paramagnetic-like regime &
ST--CF regime: many active copies, \(\hat\chi_R>0\); effective local mass
\(m_g^2=\beta_R\); no induced horizon kernel \\[8pt]

Replica-broken branch &
Ordered-like regime &
Horizon-dominated regime: \(\hat\chi_R=0\); nonlocal kernel
\(\sim \mathcal{M}^{-1}(A^h)\); induced scale
\(\gamma_{\mathrm{ind}}^4\) \\[8pt]

Order parameter &
Magnetization \(M\) &
Replica curvature \(\hat\chi_R\), which controls whether the horizon-like
kernel is generated \\[8pt]

IR scale &
\(M\) sets the correlation length &
\(\mu_{\mathrm{IR}}^2 = m^2 + \Xi_{\mathrm{rep}}\beta_R\), together with
\(\lambda^4=2g^2N\gamma_{\mathrm{eff}}^4\), determines the infrared
transverse propagator \\[8pt]

\bottomrule
\end{tabular}

\end{table*}
The analogy should not be interpreted as a literal mapping between spin
ordering and gauge fixing. Its purpose is only to emphasize that the same
replica construction can have different infrared realizations depending on the
selected branch. In the replica-symmetric branch the dominant effect is a local
CF-like screening mass, whereas in the replica-broken branch the determinant
induces a nonlocal RGZ-type horizon kernel. This is precisely why the effective
propagator can include both mechanisms in a unified notation without double
counting them as simultaneous independent contributions.
Interestingly, this statistical analogy also resonates with the structure of modern energy-based models in computational physics, suggesting that the language of spin systems may give a natural bridge between infrared gauge dynamics and optimization principles used in machine learning. A deeper exploration of this correspondeance has started in our recent work \cite{CarmoTerin:2025lyx,terin2026scaleredundancysoftgauge,CarmoTerin:2026xqu}.
The next section discusses the broader implications of this mechanism, its
limitations, and possible observables that could test the existence of an
induced horizon-like contribution.

\section{Discussion and outlook}
\label{sec:discussion}

The results of this work show that the ST replica sector contains, in the
replica-broken branch, a natural source for an RGZ-type infrared horizon kernel.
More precisely, after integrating out the replica superfields and expanding the
resulting determinant in the regulator \(\zeta\), the leading nonlocal
contribution at quadratic order in \(A^h\) has the same color and Lorentz
structure as the quadratic part of the BRST-invariant Gribov horizon functional.
This gives an induced scale
\begin{equation}
\gamma_{\mathrm{ind},R}^4
=
\kappa_R(d,N,\mu)\,\mu^2\zeta_R
+
O(\zeta_R^2),
\end{equation}
whose normalization is scheme dependent and whose sign is fixed by the
horizon-dominated infrared branch.
The conclusion should be understood in this precise sense. We have not derived
the full nonlinear horizon functional $H(A^h)$ from the ST determinant to all
orders in $A^h$. What is derived is the leading nonlocal bilinear kernel
$H_2(A^h)$, which is the part relevant for the tree-level transverse gluon
propagator. The complete RGZ horizon functional should be regarded as the
standard BRST-invariant infrared completion of this induced bilinear structure.
This distinction is important because the full horizon functional contains the
complete inverse FP operator $\mathcal M^{-1}(A^h)$, and therefore an infinite
tower of higher-order vertices in $A^h$. In the small-$\zeta$ expansion, the
higher-order terms
\[
\frac{\zeta^2}{2}\mathrm{Tr}\,\mathcal M^{-2},
\qquad
O(\zeta^3),
\]
generate additional nonlocal structures and local counterterms. These
contributions may renormalize the coefficient $\kappa_R$ and generate
subleading corrections to the effective kernel. They do not, however, modify
the leading-order statement that the linear term in $\zeta$ induces a
nonlocal bilinear structure of RGZ type.
A central point clarified by our analysis is the role of the replica phase. In
the replica-symmetric branch, $\hat\chi_R>0$, the ST determinant produces a
local CF-like screening mass. In the replica-broken branch, $\hat\chi_R=0$,
the same determinant develops nonlocal contributions involving
$\mathcal M^{-1}(A^h)$, and the induced horizon-like kernel appears. Thus the
ST sector does not simultaneously generate both the local screening mass and
the nonlocal horizon kernel within a single infrared realization.
This phase separation avoids double counting. The local mass contribution is
encoded through the factor $\Xi_{\rm rep}\beta_R$ in the replica-symmetric
branch, while the induced horizon contribution enters through
$\gamma_{\rm ind}^4$ in the replica-broken branch. The effective transverse
propagator can therefore interpolate between a CF-like form and an RGZ-like
decoupling form without adding the same infrared physics twice.

The relation between the induced scale and the usual RGZ Gribov parameter is
also clarified. In standard RGZ, the Gribov parameter is fixed by the horizon
condition. In the present mechanism, $\gamma_{\rm ind}$ is not determined by a
new independent horizon condition. Rather, it is fixed by the ST determinant
once the replica branch, the regulator $\zeta_R$, and the renormalization
scheme are specified. If an explicit RGZ horizon term is included, then the
usual horizon condition applies to the effective combination
\begin{equation}
\gamma_{\rm eff}^4
=
\gamma^4+\gamma_{\rm ind}^4.
\end{equation}
In this case the induced scale can be interpreted as a microscopic contribution
to the effective horizon strength.
The stability of the relation
$\gamma_{\mathrm{ind}}^4\propto\zeta$ under renormalization should also be
understood perturbatively in the small-$\zeta$ expansion. Renormalization
changes the coefficient $\kappa_R(d,N,\mu)$, and higher powers of
$\zeta_R$ generate additional terms,
\begin{equation}
\gamma_{\mathrm{ind},R}^4
=
\kappa_R(d,N,\mu)\,\mu^2\zeta_R
+
\kappa_R^{(2)}(d,N,\mu)\,\zeta_R^2
+
\cdots .
\end{equation}
Thus the proportionality is not meant as an exact all-order identity. It is the
leading term in the renormalized small-$\zeta$ expansion of the induced
horizon-like kernel.
The mechanism proposed here suggests several concrete observables that could be
used to test the existence of a replica-induced horizon contribution. On the
lattice, one may compare Landau-gauge ensembles obtained with different
copy-selection or copy-weighted procedures and monitor:
\begin{enumerate}
\renewcommand{\labelenumi}{(\roman{enumi})}

\item the infrared value $D_T(0)$ of the transverse gluon propagator;

\item the curvature of $D_T(p^2)$ at small momentum;

\item RGZ-inspired fits of the propagator denominator and the corresponding
effective value of $\lambda^4=2g^2N\gamma_{\rm eff}^4$;

\item the appearance or displacement of complex-conjugate poles in analytic fits
to the Euclidean propagator;

\item the behavior of ghost-sector quantities sensitive to the FP operator near
the Gribov horizon.

\end{enumerate}
In continuum approaches, the same mechanism can be tested by inserting the
induced kernel into Dyson--Schwinger or functional renormalization group
equations and studying how the infrared fixed behavior changes with
$\zeta_R$ and with the replica branch.
A complementary direction concerns analytic continuation and spectral
information. The induced horizon-like kernel gives the analytic structure of
the transverse gluon propagator and therefore provides a concrete Euclidean
input for future studies of complex singularities, positivity violation and
gauge-invariant bound-state correlators. In particular, one may investigate how
the induced scale $\gamma_{\rm ind}$ affects RGZ-type complex poles and how
these singularities reorganize inside gauge-invariant composite correlators.
Such an analysis would connect the present Euclidean construction with
spectral reconstruction, dispersive methods and possible Minkowski-space
bound-state formulations.
Several limitations should be kept in mind. First, the present derivation is
performed at leading order in the small-$\zeta$ expansion and at quadratic
order in $A^h$. Second, the coefficient $\kappa_R$ has not been computed in
a specific renormalization scheme; it has been defined through the transverse
projection of the induced kernel. Third, the existence and stability of the
replica-broken branch are assumed on the basis of the ST phase structure and
should be further tested by explicit nonperturbative calculations. Finally, a
full treatment of the higher-order terms in $\zeta$ and of the complete
operator mixing pattern remains an open problem.
Future work should therefore address:
\begin{enumerate}
\renewcommand{\labelenumi}{(\roman{enumi})}

\item the explicit evaluation of the projection coefficient
\(\kappa_R(d,N,\mu)\) in a fixed renormalization scheme;

\item the one-loop corrections to the gluon and ghost two-point functions in the
presence of the induced kernel;

\item the renormalization-group flow of \((\beta_R,\zeta_R,\gamma_{\rm eff})\);

\item the extension to linear covariant gauges, where Nielsen identities
constrain the gauge-parameter dependence of physical quantities;

\item lattice or continuum tests of the predicted phase dependence of the
infrared scales;

\item analytic-continuation studies of gauge-invariant correlators built from
the induced RGZ-type input.

\end{enumerate}
Lastly, the ST replica construction gives more than a local
copy-averaging prescription. In the replica-symmetric branch it reproduces a
CF-like screening mechanism, whereas in the replica-broken sector it induces a
nonlocal bilinear kernel of RGZ horizon type. These two mechanisms are not
competing infrared additions, but phase-dependent realizations of the same
replica framework. The result gives a microscopic and BRST-consistent origin
for the leading RGZ-type horizon kernel, and leaving the full nonlinear RGZ
horizon functional and its gap condition as the natural effective completion of
the infrared theory.

 



\appendix

\section{Superspace route from ST to a horizon kernel}
\label{sec:susy-horizon}

In this appendix we explain how the leading horizon-like bilinear kernel arises
directly from the supersymmetric formulation of the ST gauge fixing. The purpose
is to make explicit the relation between the superdeterminant of the replica
sector and the ordinary FP operator evaluated on the transverse field $A_\mu^h$.
The ST gauge fixing can be written in a topological superspace with coordinates
\[
(x^\mu,\theta,\bar\theta),
\]
where $V(x,\theta,\bar\theta)\in SU(N)$ is a group-valued nonlinear sigma
superfield and $\theta,\bar\theta$ are Grassmann coordinates. The covariant
derivative is
\begin{equation}
D_\mu V
=
\partial_\mu V+i g V A_\mu .
\end{equation}
The localized ST action has the schematic superspace form
\begin{eqnarray}
S_{\rm ST}[A,V;\beta,\zeta]
&=&
\frac{1}{2g^2}
\int d^dx\,d\theta\,d\bar\theta\,
{\rm tr}
\left[
(D_\mu V)^\dagger(D_\mu V)
+
2\zeta\,\bar\theta\theta\,
(\partial_{\bar\theta}V^\dagger)(\partial_\theta V)
\right]
\nonumber\\
&&
+
\frac{\beta}{2}\int d^dx\,A_\mu^aA_\mu^a .
\label{eq:susy-ST}
\end{eqnarray}
Introducing $n$ replicas $\{V_k\}$, integrating over the replica
superfields, and taking $n\to0$, one obtains a superdeterminant depending on
the background gauge field:
\begin{equation}
\exp[-S_{\rm rep}^{\rm eff}[A]]
\propto
\left[
{\rm sdet}
\big(\mathbb M(A)+\zeta\mathbf{1}\big)
\right]^{-1}.
\label{eq:sdet}
\end{equation}
Here $\mathbb M(A)$ denotes the superspace Hessian generated by quadratic
fluctuations of the nonlinear sigma superfield around the chosen copy.
Taking the logarithm and expanding for small $\zeta$ gives
\begin{eqnarray}
S_{\rm rep}^{\rm eff}[A]
&=&
\frac{\beta}{2}\int d^dx\,A_\mu^aA_\mu^a
-
\left[
\ln{\rm sdet}(\mathbb M+\zeta)
-
\ln{\rm sdet}\,\mathbb M
\right]
\nonumber\\
&=&
\frac{\beta}{2}\int d^dx\,A_\mu^aA_\mu^a
-
\zeta\,{\rm Str}\,\mathbb M^{-1}
+
\frac{\zeta^2}{2}\,{\rm Str}\,\mathbb M^{-2}
+
O(\zeta^3).
\label{eq:susy-expansion}
\end{eqnarray}
The supertrace in Eq.~\eqref{eq:susy-expansion} is defined by
\begin{equation}
{\rm Str}\,\mathbb M^{-1}
=
\int d^dx\,d\theta\,d\bar\theta\,
{\rm str}
\left[
\langle x,\theta,\bar\theta|
\mathbb M^{-1}
|x,\theta,\bar\theta\rangle
\right],
\label{eq:supertrace-definition}
\end{equation}
where ``str'' denotes the finite-dimensional graded trace over the internal
bosonic and fermionic components of the replica supermultiplet. The ordinary
FP operator appears after projecting the quadratic fluctuation operator onto
the physical bosonic background and performing the Grassmann integration. In
this projection, the superspace Hessian reduces to
\begin{equation}
\mathbb M(A)
\quad\longrightarrow\quad
\mathcal M(A)
=
-\partial_\mu D_\mu(A),
\label{eq:super-M-to-FP}
\end{equation}
up to local terms and BRST-exact contributions associated with the auxiliary
components of the superfield.
In the BRST-invariant formulation, the background field is replaced by the
transverse gauge-invariant field $A_\mu^h$. Therefore, for the nonlocal
gluon sector relevant to the infrared kernel,
\begin{equation}
{\rm Str}\,\mathbb M^{-1}
=
{\rm Tr}\,\mathcal M^{-1}(A^h)
+
\text{local terms}
+
\text{BRST-exact terms}.
\label{eq:Str-to-Tr}
\end{equation}
The local terms can be absorbed into local counterterms and mass
renormalizations, while the BRST-exact terms do not affect the transverse
gauge-invariant kernel considered in the main text. Hence the nonlocal part of
the supertrace is governed by the ordinary FP inverse
$\mathcal M^{-1}(A^h)$.
To extract the leading gluonic contribution, we expand
\begin{equation}
\mathcal M(A^h)
=
\mathcal M_0+V[A^h],
\qquad
\mathcal M_0^{ab}
=
-\partial^2\delta^{ab},
\qquad
V^{ab}[A^h]
=
-g f^{acb}A_\mu^{h,c}\partial_\mu .
\label{eq:susy-M0V}
\end{equation}
Then
\begin{equation}
\mathcal M^{-1}
=
\mathcal M_0^{-1}
-
\mathcal M_0^{-1}V\mathcal M_0^{-1}
+
\mathcal M_0^{-1}V\mathcal M_0^{-1}V\mathcal M_0^{-1}
+\cdots .
\label{eq:susy-Minverse}
\end{equation}
The term independent of $A^h$ contributes to the vacuum energy, and the term
linear in $A^h$ vanishes after the color trace. The leading nontrivial
contribution is therefore quadratic:
\begin{equation}
\left[
{\rm Tr}\,\mathcal M^{-1}(A^h)
\right]_{A^2}
=
{\rm Tr}
\left(
\mathcal M_0^{-1}V[A^h]\mathcal M_0^{-1}
V[A^h]\mathcal M_0^{-1}
\right).
\label{eq:susy-bilinear-trace}
\end{equation}
In coordinate space this gives
\begin{eqnarray}
\left[
{\rm Tr}\,\mathcal M^{-1}(A^h)
\right]_{A^2}
&=&
g^2
\int d^dx\,d^dy\,
A_\mu^{h,a}(x)\,
\mathcal K_{\mu\nu}^{ab}(x,y)\,
A_\nu^{h,b}(y),
\label{eq:susy-K-coordinate}
\end{eqnarray}
with
\begin{eqnarray}
\mathcal K_{\mu\nu}^{ab}(x,y)
&=&
f^{acm}f^{bdm}
\left[
\partial_\mu^x\mathcal M_0^{-1}(x-y)
\right]
\mathcal M_0^{-1}(y-x)
\left[
\partial_\nu^y\mathcal M_0^{-1}(y-x)
\right],
\label{eq:susy-kernel}
\end{eqnarray}
up to local terms and normalization conventions.
Equivalently, after projection onto the transverse part,
\begin{equation}
\left[
-\zeta\,{\rm Str}\,\mathbb M^{-1}
\right]_{A^2,\rm nonlocal}
=
\zeta g^2C_A
\int\frac{d^dp}{(2\pi)^d}\,
A_\mu^{h,a}(p)
\mathcal K_T(p^2)P_{\mu\nu}(p)
A_\nu^{h,a}(-p).
\label{eq:susy-transverse-kernel}
\end{equation}
The RGZ quadratic horizon kernel is
\begin{equation}
H_2(A^h)
=
g^2C_A
\int\frac{d^dp}{(2\pi)^d}\,
A_\mu^{h,a}(p)
\frac{P_{\mu\nu}(p)}{p^2}
A_\nu^{h,a}(-p).
\label{eq:susy-H2}
\end{equation}
Thus the transverse projection of the nonlocal part of the supertrace defines
the induced matching
\begin{equation}
\left[
-\zeta\,{\rm Str}\,\mathbb M^{-1}
\right]_{A^2,\rm nonlocal}
\equiv
\gamma_{\rm ind}^4H_2(A^h)
+
\text{local terms}
+
O((A^h)^3,\zeta^2).
\label{eq:susy-matching}
\end{equation}
The induced scale is therefore
\begin{equation}
\gamma_{\mathrm{ind},R}^4
=
\kappa_R(d,N,\mu)\,\mu^2\zeta_R
+
O(\zeta_R^2),
\label{eq:susy-gamma-ind}
\end{equation}
where the coefficient $\kappa_R$ is defined by the transverse projection of
$\mathcal K_T(p^2)$ onto the $1/p^2$ horizon structure.
This derivation makes clear why the result is a leading-order effective
correspondence. The superdeterminant produces a nonlocal bilinear kernel whose
transverse part has the RGZ horizon form. It does not, at this order, prove
that the complete nonlinear functional $H(A^h)$ is reconstructed to all
orders in $A^h$.
The induced bilinear term can be localized by introducing BRST supermultiplets
built from the Zwanziger fields,
\begin{equation}
\Phi_\mu
=
\varphi_\mu+\theta\,\omega_\mu,
\qquad
\bar\Phi_\mu
=
\bar\omega_\mu+\theta\,\bar\varphi_\mu,
\qquad
s=\partial_\theta .
\end{equation}
At quadratic order, a Hubbard--Stratonovich transformation gives
\begin{eqnarray}
\gamma_{\rm ind}^4H_2(A^h)
&\longrightarrow&
\int d^dx\,d\theta\,
\left[
\bar\Phi_\mu^{ac}\mathcal M_0^{ab}\Phi_\mu^{bc}
-
g\gamma_{\rm ind}^2 f^{abc}
A_\mu^{h,a}
(\Phi+\bar\Phi)_\mu^{bc}
\right].
\label{eq:susy-localization}
\end{eqnarray}
Expanding in components reproduces the standard Zwanziger-type localization at
quadratic order. Replacing $\mathcal M_0$ by $\mathcal M(A^h)$ gives the
usual RGZ local completion, but this completion should be regarded as the
effective BRST-invariant extension of the induced kernel.
Collecting the result, the replica-broken branch of the ST sector gives
\begin{equation}
S_{\rm rep}^{\rm eff}[A^h]
=
\frac{\beta}{2}\int d^dx\,(A_\mu^h)^2
+
\gamma_{\rm ind}^4H_2(A^h)
+
\text{local terms}
+
O((A^h)^3,\zeta^2).
\label{eq:susy-final}
\end{equation}
In the phase-dependent interpretation used in the main text, the local
$\beta$-term is active in the replica-symmetric branch, whereas the induced
horizon-like kernel is active in the replica-broken branch. This is the
superspace origin of the effective RGZ-type infrared structure discussed above.

\bibliographystyle{spphys}       
\bibliography{STGZ}   

\end{document}